\documentclass[runningheads]{llncs}
\usepackage{graphicx}
\usepackage{hyperref}
\usepackage{url}
\usepackage[utf8]{inputenc}
\usepackage{caption}
\usepackage{listings}
\usepackage{xcolor}

\definecolor{codegreen}{rgb}{0,0.6,0}
\definecolor{codegray}{rgb}{0.5,0.5,0.5}
\definecolor{codepurple}{rgb}{0.58,0,0.82}
\definecolor{backcolour}{rgb}{0.95,0.95,0.92}

\lstdefinestyle{mystyle}{
  backgroundcolor=\color{backcolour},   commentstyle=\color{codegreen},
  keywordstyle=\color{magenta},
  numberstyle=\tiny\color{codegray},
  stringstyle=\color{codepurple},
  basicstyle=\ttfamily\footnotesize,
  breakatwhitespace=false,         
  breaklines=true,                 
  captionpos=b,                    
  keepspaces=true,                 
  numbers=left,                    
  numbersep=5pt,                  
  showspaces=false,                
  showstringspaces=false,
  showtabs=false,                  
  tabsize=2
}

\makeatletter
\g@addto@macro{\UrlBreaks}{\UrlOrds}
\makeatother
\begin{document}
\title{A curated Dataset of \\ Microservices-Based Systems}
%
%
\author{Mohammad Imranur Rahman\inst{1}\orcidID{0000-0003-1430-5705} \and
Sebastiano Panichella\inst{2}\orcidID{0000-0003-4120-626X} \and
Davide Taibi\inst{1}\orcidID{0000-0002-3210-3990}}
\authorrunning{M.I. Rahman et al.}
%
\institute{CLoWEE - Cloud and Web Engineering Group. \\
Tampere University.  Tampere. 33720, Finland\\ \email{[mohammadimranur.rahman;davide.taibi]@tuni.fi}\\
\url{http://research.tuni.fi/clowee} \and
Zurich University of Applied Science (ZHAW), Zurich,  Switzerland\\
\email{panc@zhaw.ch}\\
\url{https://spanichella.github.io} 
}
\maketitle              
\begin{abstract}
Microservices based architectures are based on a set of modular, independent and fault-tolerant services.  In recent years, the software engineering community presented studies investigating potential, recurrent, effective architectural patterns in microservices-based architectures, as they are very essential to maintain and scale microservice-based systems. Indeed, the organizational structure of such systems should be reflected in so-called microservice architecture patterns, that best fit the projects and development teams needs. However, there is a lack of public repositories sharing open sources projects microservices patterns and practices, which could be beneficial for teaching purposes and future research investigations. This paper tries to fill this gap, by sharing a dataset, having a first curated list microservice-based projects. Specifically, the dataset is composed of 20 open-source projects, all using specific microservice architecture patterns.  Moreover, the dataset also reports information about inter-service calls or dependencies of the aforementioned projects. For the analysis, we used two different tools (1) SLOCcount and (2) MicroDepGraph to get different parameters for the microservice dataset. Both the microservice dataset and analysis tool are publicly available online.
We believe that this dataset will be highly used by the research community for understanding more about microservices architectural and dependencies patterns, enabling researchers to compare results on common projects.

\end{abstract}
\section{Introduction}

Microservices based architectures are based on a set of modular, independent and fault-tolerant services, which are  ideally easy to be monitored and tested \cite{MartinP19}, and can be easily maintained~\cite{TaibiIEEECloud} by integrating also user feedback in the loop \cite{PanichellaSGVCG15,GranoCPPG18}. However, in practice, decomposing a monolithic system into independent microservices is not a trivial task \cite{TaibiXP17}, which is typically performed manually by software architects ~\cite{TaibiIEEECloud,SOLDANI2018}, without the support of  tool automating  the decomposition or slicing phase~\cite{TaibiIEEECloud}.
To ease the identification of microservices in monolithic applications, further empirical investigations need to be performed and automated tools (e.g., based on summarization techniques \cite{Panichella18}) need to be provided to developers, to make this process more reliable and effective~\cite{TaibiCLOSER}.

In recent years, the software engineering community presented studies investigating the potential, recurrent, effective architectural patterns~\cite{TaibiIEEECloud,TaibiCLOSER} and anti-patterns~\cite{TaibiIEEEsw,TaibiBOOK,TaibiBOOK1} in microservices-based architectures. Indeed, the organizational structure of such systems should be reflected in so-called microservice architecture patterns, that best fit the projects and development teams needs. However, there is a lack of public repositories sharing open sources projects microservices patterns and practices, which could be beneficial for teaching purposes and future research investigations.

This paper tries to fill this gap, by sharing a dataset, having a first curated list of open-source microservice-based projects. Specifically, the dataset is composed of 20 open-source projects, all using specific microservice architecture patterns.  Moreover, the dataset also reports information about inter-service calls or dependencies of the aforementioned projects. For the analysis, we used two different tools such as (1) SLOCcount and (2) MicroDepGraph. The microservice dataset~\cite{MSDataset2019} and analysis tool~\cite{microdepgraph} are publicly available online, and detailed in the following sections.

At the best of our knowledge only Márquez and Hastudillo proposed a dataset of microservices-based projects~\cite{Marquez}. However, their goal was the investigation of architectural patterns adopted by the microservices-based projects, and they did not provided dependency graphs of the services. 

We believe that this dataset will be highly used by the research community for understanding more about microservices architectural and dependencies patterns, enabling researchers to compare results on common projects.

\textbf{Paper structure}. In Section \ref{sec:background}, we discuss the main background of this work, focusing on the open challenges concerning understanding an analyzing microservices-based architectures. In Section \ref{sec:prjselection}, we discuss the projects selection strategy, while in Section\ref{sec:datacollection} are described the data extraction process (describing the tools used and implemented for it) and the generated data. Finally, Section \ref{sec:threats} and Section \ref{sec:conclusion}, discuss the main threats of concerning the generation of the generated dataset, concluding the paper outline future directions. 
 
\section{Background}
\label{sec:background}

In recent years, the software industry especially the enterprise software are rapidly adopting the Microservice architectural pattern. Compared to a service-oriented architecture, the microservice architecture is more decoupled, independently deployable and also horizontally scalable. In microservices, each service can be developed using different languages and frameworks. Each service is deployed to their dedicated environment whatever efficient for them. The communication between the services can be either REST or RPC calls. So that whenever there is a change in business logic in any of the services others are not affected as long as the communication endpoint is not changed. As a result, if any of the components of the  system fails, it will not affect the other components or services, which is a big drawback of monolithic system~\cite{Fowler2014}. The clear separation of tasks between teams developing microservices also enable teams to deploy independently. Another benefit of microservices is that the usage of DevOps is simplifies~\cite{TaibiMSDEvOps}. The drawback, is the increased initial development effort, due to the connection between services~\cite{Nyyti19}.

\begin{figure*}[!h]
\centering 
\includegraphics[trim= 0 50 0 0,  clip, width=1\linewidth]{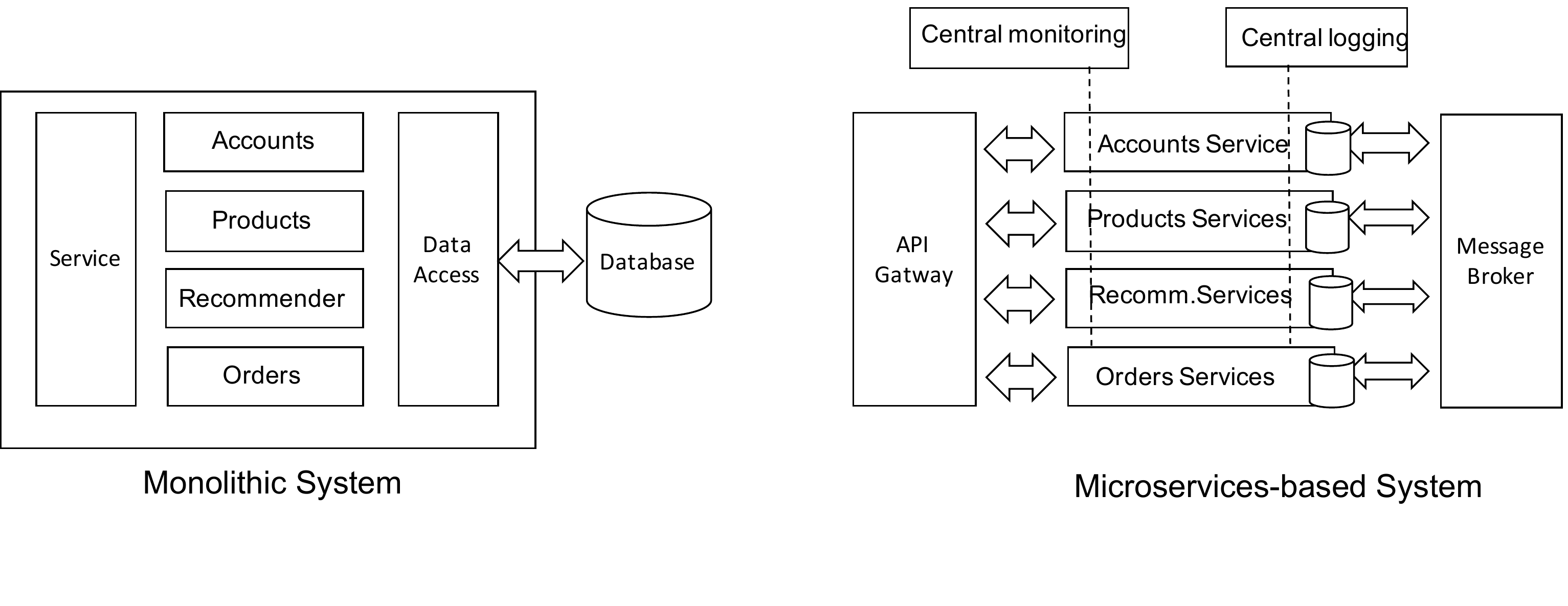}
\caption{Architectures of Monolithic and Microservices systems}
\label{fig:Microservice}
\end{figure*}

As  we can see in Figure \ref{fig:Microservice},  components in monolithic systems are tightly coupled with each other so that the failure of one component will affect the whole system. Also if there are any architectural changes in a monolithic system it will affect other components. Due to these advantages, microservice architecture is way more effective and efficient than monolithic systems. Instead of having lots of good features of microservice, implementing and managing microservice systems are still challenging and require highly skilled developers~\cite{Balalaie2016}.

\section{Project Selection}
\label{sec:prjselection}
We selected projects from GitHub, searching projects implemented with a microservice-based architecture, developed in Java and using docker. 

The search process was performed applying the following search string: 
\begin{verbatim}
"micro-service" OR microservice OR "micro-service" 
filename:Dockerfile language:Java
\end{verbatim}

\begin{table}[!h]
\centering
\caption{The projects in the dataset}
\label{tab:projects}
\resizebox{\textwidth}{!}{%
\begin{tabular}{|l|l|c|c|c|c|c|}
\hline
\textbf{Project Name}                             & \textbf{Project Repository}    & \textbf{\#Ms.} & \textbf{KLOC}   & \textbf{\#Commits} & \textbf{\#Dep.} & \textbf{Project Type} \\ \hline
Consul demo                              & http://bit.ly/2KsGzx6 & 5               & 2.343  & 78        & 4              &      Demo            \\
CQRS microservice application            & http://bit.ly/2YtbtiF & 7               & 1.632  & 86        & 3              &      Demo            \\
E-Commerce App                           & http://bit.ly/2yLqTPW & 7               & 0.967  & 20        & 4              &      Demo            \\
EnterprisePlanner                        & http://bit.ly/2ZPK7je & 5               & 4.264  & 49        & 2              &      Demo            \\
eShopOnContainers                        & http://bit.ly/2YGSkJB & 25              & 69.874 & 3246      & 18             &      Demo            \\
FTGO - Restaurant Management             & http://bit.ly/2M7f8fm & 13              & 9.366  & 172       & 9              &      Demo            \\
Lakeside Mutual Insurance Company        & http://bit.ly/33iJSiU & 8               & 19.363 & 12        & 7              &      Demo            \\
Lelylan - Open Source Internet of Things & http://bit.ly/2TdDfd3 & 14              & 77.63  & 2059      & 11             &                  Industrial      \\
Microservice Architecture for blog post  & http://bit.ly/2OKY29v & 9               & 1.536  & 90        & 7              &      Demo            \\
Microservices book                       & http://bit.ly/2TeSbI2 & 6               & 2.417  & 127       & 5              &      Demo            \\
Open-loyalty                             & http://bit.ly/2ZApXtA & 5               & 16.641 & 71        & 2              &                  Industrial      \\
Pitstop - Garage Management System       & http://bit.ly/2Td7NLY & 13              & 34.625 & 198       & 9              &      Demo            \\
Robot Shop                               & http://bit.ly/2ZFbHQm & 12              & 2.523  & 208       & 8              &      Demo            \\
Share bike (Chinese)                     & http://bit.ly/2YMJgmb & 9               & 3.02   & 62        & 6              &      Demo            \\
Spinnaker                                & http://bit.ly/2YQA2S7 & 10              & 33.822 & 1669      & 6              &          Industrial        \\
Spring Cloud Microservice Example        & http://bit.ly/2GS2ywt & 10              & 2.333  & 35        & 9              &      Demo            \\
Spring PetClinic                         & http://bit.ly/2YMVbAC & 8               & 2.475  & 658       & 7              &      Demo            \\
Spring-cloud-netflix-example             & http://bit.ly/2YOUJxJ & 9               & 0.419  & 61        & 6              &      Demo            \\
Tap-And-Eat (Spring Cloud)               & http://bit.ly/2yIjXmC & 5               & 1.418  & 35        & 4              &      Demo            \\
Vehicle tracking                         & http://bit.ly/31i5aLM & 8               & 5.462  & 116       & 5              &      Demo\\
\hline
\end{tabular}
}
\end{table}

Results of this query reported 18,639 repository results mentioning these keywords. 

We manually analyzed the first 1000 repositories, selecting projects implemented with a microservice-architectural style and excluding libraries, tools to support the development including frameworks, databases, and others. 

Then, we created a github page to report the project list~\cite{MSDataset2019}
and we opened several questions on different forums\footnote{ResearchGate. \url{https://www.researchgate.net/post/Do_you_know_any_Open_Source_project_that_migrated_form_a_monolithic_architecture_to_microservices}} 
\footnote{Stack Overflow -1  \url{https://stackoverflow.com/questions/48802787/open-source-projects-that-migrated-to-microservices}}. Moreover, we monitored replies to similar questions on other practitioners forums\footnote{Stack Overflow -2 \url{https://stackoverflow.com/questions/37711051/example-open-source-microservices-applications}} 
\footnote{Stack Overflow -3 \url{https://www.quora.com/Are-there-any-examples-of-open-source-projects-which-follow-a-microservice-architecture-DevOps-model}} \footnote{Quora -1 \url{https://www.quora.com/Are-there-any-open-source-projects-on-GitHub-for-me-to-learn-building-large-scale-microservices-architecture-and-production-deployment}} 
\footnote{Quora -2  \url{https://www.quora.com/Can-you-provide-an-example-of-a-system-designed-with-a-microservice-architecture-Preferably-open-source-so-that-I-can-see-the-details}}
to ask practitioners if they were aware of other relevant Open Source projects implemented with a microservice-architectural style.  
We received 19 replies from the practitioners' forums, recommending to add 6 projects to the list. 
Moreover, four contributors send a pull request to the repository to integrate more projects. 

In this work, we selected the top 20 repositories that fulfill our requirements. 

The complete list of projects is available in Table~\ref{tab:projects} and can be downloaded from the repository GitHub page~\cite{microdepgraph}. 

\section{Data Collection}
\label{sec:datacollection}
We analyzed different aspects of the projects. 
We first considered the size of the systems, analyzing the size of each microservices in Lines of code. The analysis was performed by applying the SLOCCount tool\footnote{SLOCcount. https://dwheeler.com/sloccount/}. 

Then we analyzed the dependencies between services by applying the MicroDepGraph tool~\cite{microdepgraph} developed by one of the authors. 

\subsection{SLOCcount}

SLOCcount is an open source tool for counting the effective lines of code of an application. It can be executed on several development languages, and enable to quickly count the lines of code in different directories.

\subsection{MicroDepGraph}
MicroDepGraph is our in-house tool developed for detecting dependencies and plot the dependency graph of microservices. 

Starting from the source code of the different microservices, it analyzes the docker files for service dependencies defined in docker-compose and java source code for internal API calls. The tool is completely written in Java. It takes two parameters as input: (1) the path of the project in the local disk and (2) the name of the project

We chose to analyze docker-compose files because, in microservices projects, the dependencies of the services are described in the docker-compose file as configuration. As the docker-compose is a YML or YAML file so the tool parses the files from the projects. MicroDepGraph first determines the services of the microservices project defined in the docker-compose file. Then for each service, it checks dependencies and maps the dependencies for respective services.

Analyzing only the docker-compose file does not give us all relationships of the dependencies, as there might be internal API call, for example, using a REST client. For this reason, we had to analyze the java source code for possible API calls to other services. As we are analyzing Java microservices project, the most commonly used and popular framework for building microservices in java is Spring Boot. In spring boot the API endpoints for services are configured and defined using different annotations in java source code. So we targeted these annotations when parsing java source code. First, we determined the endpoints for each service by parsing the java source code and looking for the annotations that define the endpoints. For parsing Java source code we used an open source library called JavaParser\footnote{JavaParser. https://javaparser.org/}. After getting endpoints for each service we searched whether there are any API calls made from other services using these endpoints. Then if there is an API call of one service from another, we map it as a dependency and add it to our final graph. After finding all the mapping the tool then makes relationships(dependencies) between the services and draws a directed graph. 

Finally, it generates a graph representation formatted as GraphML file, a neo4j database containing all the relationships and an svg file containing the graph.

\begin{figure}
    \centering
    \includegraphics[width=8cm]{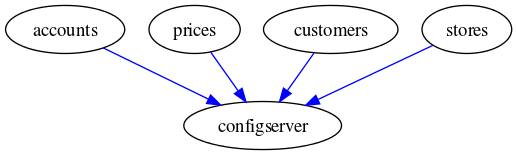}
    \caption{Dependency graph}
    \label{fig:depgraph}
\end{figure}

Figure~\ref{fig:depgraph} shows an example of the output provided by MicroDepGraph on the project ''Tap And Eat''.

\section{Dataset production and Structure}
For each project, we first cloned the repository. Then we executed SLOCcount independently on each project to extract the number of lines of code. 
Then we executed MicroDepGraph to obtain the dependencies between the microservices. From MicroDepGraph we got GraphML and svg file for each project. To generate GraphML file we used Apache TinkerPop\footnote{Apache TinkerPop  http://tinkerpop.apache.org/} graph computing framework. The GraphML file is easy to use xml based file where we can specify directed or undirected graphs and different attributes to the graph. Moreover, we can import the GraphML file in different graph visualization platforms like Gephi\footnote{Gephi  https://gephi.org/}. In this kind of graph visualization tools we can then apply different graph algorithms for further analyzing the graph. We also get SVG image as output so that it can be easily used for further processing.

Finally, we stored the results in a Github repository~\cite{MSDataset2019}  as graphml files, together with the list of analyzed microservice projects. Below is an output of one of the projects analyzed by MicroDepGraph including the GraphML output,

{}


\label{lst}
\begin{lstlisting}[language=Xml, caption=GraphML file , frame=lines]
<?xml version="1.0" encoding="UTF-8"?>
<graphml xmlns="http://graphml.graphdrawing.org/xmlns" xmlns:xsi="http://www.w3.org/2001/XMLSchema-instance" xsi:schemaLocation="http://graphml.graphdrawing.org/xmlns http://graphml.graphdrawing.org/xmlns/1.1/graphml.xsd">
   <key id="edgelabel" for="edge" attr.name="edgelabel" attr.type="string" />
   <graph id="G" edgedefault="directed">
      <node id="stores" />
      <node id="configserver" />
      <node id="accounts" />
      <node id="customers" />
      <node id="prices" />
      <edge id="stores-&gt;configserver" source="stores" target="configserver" label="depends">
         <data key="edgelabel">depends</data>
      </edge>
      <edge id="accounts-&gt;configserver" source="accounts" target="configserver" label="depends">
         <data key="edgelabel">depends</data>
      </edge>
      <edge id="customers-&gt;configserver" source="customers" target="configserver" label="depends">
         <data key="edgelabel">depends</data>
      </edge>
      <edge id="prices-&gt;configserver" source="prices" target="configserver" label="depends">
         <data key="edgelabel">depends</data>
      </edge>
   </graph>
</graphml>
\end{lstlisting}






\textbf{License}.
The dataset has been developed only for research purposes. It includes data elaborated and extracted from public repositories.  Information from GitHub is stored under GitHub Terms of Service (GHTS), which explicitly allow extracting and redistributing public information for research purposes\footnote{GitHub Terms of Service. goo.gl/yeZh1E  Accessed: July 2019}. 

The dataset is licensed under a Creative Commons Attribution-NonCommercial- ShareAlike 4.0 International license.

\section{Threats to Validity}
\label{sec:threats}
We are aware that both SLOCcount and MicroDepGraph  might  analyze the projects incorrectly under some conditions. 
Moreover, regarding SLOCcount we analyzed only the Java lines of code. We are aware that some project could contain also code written in other language or that the tool  could provide  incorrect results. 

Another important threat is related to the generalization of the dataset. We selected the list of projects based on different criteria (see Section~\ref{sec:prjselection}). Moreover, several projects are toy-projects or teaching examples and they cannot possibly represent the whole open-source ecosystem. Moreover, since the dataset does not include industrial projects, we cannot make any speculation on closed-source projects.

\section{Conclusion}
\label{sec:conclusion}

In this paper, we presented a curated dataset for microservices-based systems. To analyze the microservices projects we developed a tool(MicroDepGraph) to determine the dependencies of services in the microservices project. 

We analyzed 20 open source microservice projects which include both demo and industrial projects. The number of services in the projects ranges from 5 to 25. To analyze dependencies we considered docker and internal API calls. Due to the docker analysis, this tool can analyze any microservice system that uses docker environment regardless of programming languages or frameworks. But for the API call, it will only analyze the projects implemented using Spring framework. As Spring framework is widely used for developing microservice systems.

The output of the tool will allow researchers as well as companies to analyze the dependencies between each service in microservice project so that they can improve the architecture of the system. The output contains both GraphML and SVG file for further analysis. 

We are planning to extend the tool so that it can analyze microservices developed in any framework and programming language. Also, we can use machine learning approach to identify different parameters and anomalies in the architecture of microservice systems.
Moreover, we are planning to calculate quality metrics for microservices, based on~\cite{TaibiSysta19,Bogner2017}

%
%
%
\bibliographystyle{splncs04}
\bibliography{mybibliography}
%




\end{document}